# A Robust Target Linearly Constrained Minimum Variance Beamformer With Spatial Cues Preservation for Binaural Hearing Aids

Hala As'ad , Martin Bouchard , and Homayoun Kamkar-Parsi

*Abstract*—In this paper, a binaural beamforming algorithm for hearing aid applications is introduced. The beamforming algorithm is designed to be robust to some error in the estimate of the target speaker direction. The algorithm has two main components: a robust target linearly constrained minimum variance (TLCMV) algorithm based on imposing two constraints around the estimated direction of the target signal, and a post-processor to help with the preservation of binaural cues. The robust TLCMV provides a good level of noise reduction and low level of target distortion under realistic conditions. The post-processor enhances the beamformer abilities to preserve the binaural cues for both diffuse-like background noise and directional interferers (competing speakers), while keeping a good level of noise reduction. The introduced algorithm does not require knowledge or estimation of the directional interferers' directions nor the second-order statistics of noise-only components. The introduced algorithm requires an estimate of the target speaker direction, but it is designed to be robust to some deviation from the estimated direction. Compared with recently proposed state-of-the-art methods, comprehensive evaluations are performed under complex realistic acoustic scenarios generated in both anechoic and mildly reverberant environments, considering a mismatch between estimated and true sources direction of arrival. Mismatch between the anechoic propagation models used for the design of the beamformers and the mildly reverberant propagation models used to generate the simulated directional signals is also considered. The results illustrate the robustness of the proposed algorithm to such mismatches.

## I. INTRODUCTION

A HEARING aid is a common and effective solution to sensorineural hearing loss. Despite enormous advances in hearing aid technology, the performance of hearing aids under noisy environments remains one of the most common complaints from hearing aid users [1], [2], and hearing-impaired people face challenges in understanding and separating speech in noisy environments [1]–[3].

For noise reduction, single channel processing algorithms, which rely on frequency and temporal information of the input signals, have been extensively researched such as in [4], [5]. However, single channel algorithms suffer from several limitations under low-SNR acoustic scenarios, especially for non-stationary noise and multi-talkers conditions. Single channel solutions typically also introduce distortion and do not provide true speech intelligibility improvement. A notable exception is the solution in [6] which has been found to improve speech intelligibility. The solution in [6] is based on deep neural networks and a binary masking of some speech components in the T-F domain. This solution, however, does not preserve naturalness of the target speaker speech (high distortion), which is a concern for its use in hearing aids. It has also not been developed for the case of one or two competing talkers.

As an alternative, microphone array processing (beamforming) has been widely used in modern hearing aids, leading to directionally sensitive hearing aids [7]. Binaural hearing aids have also recently been introduced in the market. Binaural hearing aids have a hearing aid device at each ear, each possibly equipped with multiple microphones, and the devices are capable to transmit signals or information from one side to the other through a "binaural wireless link". Microphone arrays can provide good noise reduction with low distortion, and the use of additional microphones and different microphone geometry in binaural hearing aids can lead to further improvements in the directional response, compared to monaural single-sided beamforming. However, even binaural hearing aids have still not achieved the required robustness in case of real-life complex environments [8]. The performance of binaural beamformers can be significantly affected by a mismatch or an error between the target source propagation model assumed for the beamformer design and the actual physical target source propagation [9], [10]. This includes errors in the estimated target direction of arrival (DOA) used in the beamformer algorithms, i.e., target DOA mismatch. This kind of mismatch can be generated from imperfect target DOA estimation schemes, from small head movements of the hearing aid user, and from multipath propagation. To address this problem, several acoustic beamforming methods robust to the mismatch in target propagation models have been introduced in the literature [11]–[21], and some of these solutions are not specifically for binaural hearing aids.

Manuscript received December 19, 2018; revised April 24, 2019 and June 16, 2019; accepted June 17, 2019. Date of publication June 21, 2019; date of current version July 1, 2019. This work was supported in part by a Natural Sciences and Engineering Research Council Discovery grant. The associate editor coordinating the review of this manuscript and approving it for publication was Prof. Simon Doclo. *(Corresponding author: Martin Bouchard.)*

H. As'ad and M. Bouchard are with the School of Electrical Engineering and Computer Science, University of Ottawa, Ottawa, ON K1N 6N5, Canada (e-mail: hasad056@uottawa.ca; martin.bouchard@uottawa.ca).

H. Kamkar-Parsi is with WS Audiology 91058, Erlangen, Germany (e-mail: homayoun.kamkarparsi@sivantos.com).

Digital Object Identifier 10.1109/TASLP.2019.2924321

Unfortunately, most of the previous work rely on sophisticated Voice Activity Detection (VAD), speech presence probability estimation, and/or SNR estimation. These can become difficult to measure in complicated multi-talker reverberant environments, with speakers having variable activity patterns. An interesting solution for hearing aids based on inequality constrained optimization has been proposed in [22] and discussed in [23], to increase the robustness to target DOA mismatch. However, since this design uses extra constraints for directional sources to increase robustness to DOA mismatch, this can lead to low degrees of freedom available for residual noise reduction (e.g., low number of adaptive "nulls") in case of limited number of available microphones signals. In addition, it requires an estimation of the DOA for the directional interferer sources.

All the beamforming designs in [11]–[21] were not designed to preserve the binaural cues of the residual directional interferers and diffuse-like noise in the binaural output signals. Several binaural beamforming solutions have been introduced to preserve some of the binaural cues of these components, while also preserving the target signal and achieving a good noise reduction level. Under some assumptions (e.g., accurate direction of arrival estimates), binaural beamforming processing such as the second and third methods in [24] can provide directional noise reduction and preserve the binaural cues of the target signal and the directional interferers, depending on the number of available microphones. However, this binaural beamforming is not designed to preserve the binaural cues of the diffuse-like background noise. The Multichannel Wiener Filter (MWF) is the basis of several proposed solutions that aim to preserve the binaural cues. Extensions of the MWF have been proposed in [25]–[29] as attempts to preserve the binaural cues for the different acoustic scene components. A potential challenge for the MWF and its extensions is the need for an accurate estimate of the second order statistics for the noise-only components, which can be difficult to achieve in complex acoustic environments, for example multiple talkers with time-varying activity patterns and statistics. Detailed information of the MWF and its extensions can be found in [30].

The Binaural Linearly Constrained Minimum Variance method (BLCMV) has been introduced in [31] and a comprehensive theoretical analysis has been provided in [32]. The BLCMV is capable to provide a good trade-off between noise reduction and cues preservations for a limited number of interferer sources. As an attempt to enhance the noise reduction abilities of the BLCMV, an optimal BLCMV has also been proposed in [33]. However, the optimal BLCMV is capable to preserve the binaural cues for just one directional interferer as well as the target source. As another variation of the BLCMV, joint BLCMV, which jointly estimates the left and right beamformers of two hearing aids, has been introduced in [34] in order to enhance the binaural cues preservations abilities of the BLCMV. The joint BLCMV needs one constraint per interferer to preserve the binaural cues of the interferers, unlike the BLCMV which uses two constraints to preserve each interferer. However, since a limited number of microphones are available in binaural hearing aids, the joint BLCMV can still face a degradation of performance when the number of sources increases. A relaxed version of the joint BLCMV has been proposed in [35]. In this relaxed BLCMV, tunable parameters have been used for each directional interferer, in order to separately control the trade-off between the binaural cues preservation and the noise reduction for each interferer. The BLCMV and all its extensions, i.e., [31]-[35], require knowledge of the propagation models for the directional interferers in addition to the target source (directivity vectors, steering vectors, Relative Acoustic Transfer Functions (RATF)). As will be shown in this paper, this can limit the performance of these approaches, as they suffer from errors in the estimated propagation models. In addition, the BLCMV and its variations do not have the ability to preserve the binaural cues of the diffuse-like background noise. As an attempt to design a BLCMV beamformer that does not depend on the propagation models (and directions of arrival) of the directional sources, a set of pre-determined RATFs distributed around the head have been used for beamforming design in [36]. Each RATF is responsible for preserving the binaural cues of the directional sources coming from certain directions. Increasing the number of predetermined RATFs decreases the effect of the mismatch between the true and the pre-determined RATFs, but it also requires a larger number of microphones in order to achieve a good performance. However, in hearing aids applications, only a small number of microphones are normally available for the binaural beamformer.

In order to preserve the binaural cues for directional interferers and diffuse-like noise components without a knowledge of the propagation model of the directional interferers, a binary decision/classifier algorithm common to the left and right beamformer outputs for each time-frequency (T-F) bin was proposed in [37], [38]. A challenge for this classification algorithm is its applicability in low input SNR environments, as most T-F bins can be classified as noise-dominant, resulting in low SNR improvement and an attenuated target output, as illustrated in [39]. As an attempt to enhance the performance of this method, the classification mechanism was later modified to use the output SNR instead of the input SNR [40]. However, this method requires an estimation of the second order statistics of the noise and the target components, which, as previously described, can be challenging in some real-life time-varying multi-talker environments. In our recent work [41], an algorithm based on classification and mixing of binaural signals at each T-F bin was introduced. Three classification criteria were proposed, based on the power, power difference, and complex coherence computed from: 1) binaural beamformer output signals (with good level of noise reduction) and 2), original binaural noisy signals (or alternatively, other binaural signals with cues preserved but with an intermediate level of noise reduction [42], [43]). The complex coherence criterion provided better noise reduction over the other classification criteria.

In this work, we contribute in 1) designing a binaural beamformer which is robust to mismatch in target propagation models, 2) proposing a modified post-processor method preserving the binaural cues of all acoustic scene components (target, diffuse-like background, directional interferers), with a good tradeoff between noise reduction and cues preservations. For the first contribution, we introduce the Robust TLCMV which is robust to a mismatch (error) in the directivity vector assumed for the target signal. This is achieved by designing a binaural beamformer with a wider beam around the estimated target direction. For the second contribution, the binaural cues preservation are achieved by using a simplified and improved version of the coherence-based post-processor method in [41], for classification and mixing of binaural signals. Both the proposed Robust TLCMV and the post-processor do not rely on any assumption for the propagation model (or DOAs) of the interferers (competing speakers). The proposed TLCMV with post-processor is also found to be robust to both target DOA mismatch and mismatch between the anechoic propagation model used for the beamformer design and the mildly reverberant propagation models used to generate the directional signals in the simulations. The proposed solution does not rely on target VAD detection, speech probability presence estimation, or SNR estimation, which can be difficult to compute in complex real-life time-varying multi-talker environments.

In order to study the robustness of the proposed algorithm to different types of mismatches, comprehensive validations are conducted through simulations using acoustic scenarios generated in mildly reverberant environments and anechoic environments (i.e., with and without mismatch in the sets of directivity/steering vectors), and for scenarios with and without DOA mismatch. Comparisons are performed with the recently proposed state-of-the-art Binaural Minimum Variance Distortionless Response (BMVDR) beamformer, the BMVDR with partial noise estimation (BMVDR-n) beamformer [29], [44], [45] and with the BLCMV beamformers which uses constraints to attenuate interferers.

This paper is organized as the following. Section II provides a detailed description of the system notations and the beamforming microphone configurations that are used throughout this paper. Section III provides a summary of the previously proposed BLCMV, BMVDR and BMVDR with partial noise estimation algorithms. Section IV provides some detailed information about the new beamforming algorithm and the post-processing algorithm proposed in this work. Section V explains the performance metrics used in this work. Section VI explains the experimental setup. Finally, Section VII provides the simulation results of the proposed algorithms, and performance comparisons with the state-of-the-art algorithms.

## II. SYSTEM NOTATIONS AND REFERENCE BEAMFORMING PROCESS

### A. System Notations

Binaural hearing aid units with two microphone arrays of $M/2$ microphones at each ear, i.e., $M$ microphones in total, and ideal binaural wireless links between the units (no jitter, delay, packet loss, etc.) are considered. A Short Time Fourier Transform (STFT) is used in order to represent the input signals in the Time-Frequency (T-F) domain. The input noisy microphone signals in the T-F domain can be written as in (1), with the microphone signals transmitted from one side to the other side through the binaural wireless links:

$$y_m(f,t) = x_{in,m}(f,t) + v_{in,m}(f,t) + n_{in,m}(f,t) \quad (1)$$

where $m$ is the microphone index, and

$$m = \underbrace{1,...,M/2}_{\text{left side}}, \underbrace{M/2+1,...,M}_{\text{right side}}.$$

The front left (FL) microphone has index $m = 1$, and the front right (FR) microphone has index $m = M/2 + 1$. These microphones are the reference microphone for the left-side beamformer and the right-side beamformer, respectively. $x_{in,m}$, $v_{in,m}$, and $n_{in,m}$ at the $m$th microphone are the target speaker, the sum of directional interferer speakers, and the diffuse-like background noise components, respectively. $f$ is the frequency index and $t$ is the time (frame) index.

By stacking the input microphone signals in $M$ dimensional vectors, the input signals from the left and right microphones can be written as in (2):

$$\mathbf{y}(f,t) = \mathbf{x}(f,t) + \mathbf{v}(f,t) + \mathbf{n}(f,t) \quad (2)$$

where, $\mathbf{y}(f,t) = [y_1(f,t), y_2(f,t), ..., y_M(f,t)]^T$

$$\mathbf{x}(f,t) = [x_{in,1}(f,t), x_{in,2}(f,t), ..., x_{in,M}(f,t)]^T$$

$$\mathbf{v}(f,t) = [v_{in,1}(f,t), v_{in,2}(f,t), ..., v_{in,M}(f,t)]^T$$

$$\mathbf{n}(f,t) = [n_{in,1}(f,t), n_{in,2}(f,t), ..., n_{in,M}(f,t)]^T.$$

Assuming that $s_x$ is a target source signal coming from angle $\theta_x$, and $s_{vi}$ is the $i$th interferer source signal (competing talker) coming from angle $\theta_{vi}$, the target component and the sum of the directional interferer components at the microphones can be written in terms of the directivity vectors $\mathbf{d}(f, \theta)$ as in (3) and (4), respectively:

$$\mathbf{x}(f,t) = \mathbf{d}(f, \theta_x) s_x(f,t) \quad (3)$$

$$\mathbf{v}(f,t) = \sum_{i=1}^{N} \mathbf{d}(f, \theta_{vi}) s_{vi}(f,t). \quad (4)$$

The vector $\mathbf{d}(f, \theta_x) = [d_1(f, \theta_x), ..., d_M(f, \theta_x)]^T$ is the target directivity vector, which is the frequency response between the target source and each microphone. Likewise, the vector $\mathbf{d}(f, \theta_{vi}) = [d_1(f, \theta_{vi}), ..., d_M(f, \theta_{vi})]^T$ is the interference directivity vector for source $s_{vi}$, which is the frequency response between the interference source $s_{vi}$ and each microphone. $N$ is the number of directional interferers in the acoustic scenario considered. In hearing aids, the directivity vectors include the head shadow effect and other head/ear related effects (e.g., pinnae filtering), therefore Head-Related Transfer Functions (HRTFs) are used for the directivity vectors in the beamformer designs.

The input target signal at the reference microphone $x_{ref}(f,t)$ can be defined as in (5):

$$x_{ref}(f,t) = d_{ref}(f, \theta_x) s_x(f,t). \quad (5)$$

If the reference microphone is the FL, then $x_{ref}(f,t) = x_{in,1}(f,t)$. If the reference microphone is the FR, then

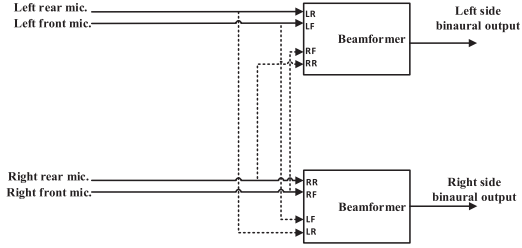

Fig. 1.   2 + 2 microphone configuration (dotted lines represent signals transmitted through a wireless link).

$x_{ref}(f,t) = x_{in,M/2+1}(f,t)$. Likewise, $d_{ref}(f,\theta_x)$ is the target directivity vector (or HRTF) at the reference microphone, with $d_{ref}(f,\theta_x) = d_1(f,\theta_x)$ if FL, and $d_{ref}(f,\theta_x) = d_{M/2+1}(f,\theta_x)$ if FR.

A correlation matrix for the target component can be defined as in (6):

$$
\begin{aligned}
\mathbf{R}_x(f) &= E\{\mathbf{x}(f,t)\mathbf{x}^H(f,t)\} \\
&= E\{\mathbf{d}(f,\theta_x)s_x(f,t)\mathbf{d}^H(f,\theta_x)s_x^*(f,t)\} \\
&= \mathbf{d}(f,\theta_x)\mathbf{d}^H(f,\theta_x)E\{|s_x(f,t)|^2\}.
\end{aligned}
\tag{6}
$$

The superscript $H$ refers to "Hermitian" which is the complex conjugate transpose, the superscript "*" refers to the complex conjugate, and $E\{.\}$ refers to the expectation operator. Similarly, the correlation matrix of the sum of directional interferer components can be defined as in (7) and (8). Directional interferers are assumed to be uncorrelated with each other.

$$
\begin{aligned}
\mathbf{R}_v(f) &= E\{\mathbf{v}(f,t)\mathbf{v}^H(f,t)\} \\
&= E\left\{\left(\sum_{i=1}^N \mathbf{d}(f,\theta_{vi})s_{vi}(f,t)\right)\left(\sum_{i=1}^N \mathbf{d}(f,\theta_{vi})s_{vi}(f,t)\right)^H\right\} \\
&= \sum_{i=1}^N \mathbf{d}(f,\theta_{vi})\mathbf{d}^H(f,\theta_{vi})E\{|s_{vi}(f,t)|^2\}
\end{aligned}
\tag{7}
$$

The correlation matrix of the diffuse-like background noise component is defined as in (8):

$$
\mathbf{R}_n(f) = E\{\mathbf{n}(f,t)\mathbf{n}^H(f,t)\}.
\tag{8}
$$

Assuming that the target component, the sum of directional interferer components, and the diffuse-like background noise component are uncorrelated, the correlation matrix of the input noisy signals can be written as in (9):

$$
\mathbf{R}_y(f) = \mathbf{R}_x(f) + \mathbf{R}_v(f) + \mathbf{R}_n(f).
\tag{9}
$$

### B. Beamformer Microphone Configuration

In this work, a binaural hearing aid with two microphones on each side of the head is used, as illustrated in Fig. 1. We take advantage of the availability of two bidirectional binaural wireless links to transmit two microphone signals from each side to the other side. Thus, the beamformer on each side has direct access to four microphone signals. We will refer to this design as the 2 + 2 microphone configuration. The binaural beamformers are used to process the input noisy signals as in

(10) and (11), to generate the left and right beamformer outputs $z_l(f,t)$ and $z_r(f,t)$, respectively. The binaural beamformer on the left side aims to extract the target signal as received at the FL microphone (i.e., using the FL microphone as a reference microphone). The binaural beamformer on the right side aims to extract the target signal as received at the FR microphone (i.e., using the FR microphone as a reference microphone).

$$
z_l(f,t) = \mathbf{w}_l^H(f,t)\mathbf{y}(f,t)
\tag{10}
$$

$$
z_r(f,t) = \mathbf{w}_r^H(f,t)\mathbf{y}(f,t)
\tag{11}
$$

## III. REVIEW OF PREVIOUS BINAURAL BEAMFORMING ALGORITHMS

In this section, we review the BLCMV, the BMVDR (which is a special case of the BLCMV) and the BMVDR extension with partial noise estimation (BMVDR-n).

### A. Binaural LCMV (BLCMV)

The BLCMV [31], [32] is a general form of the BMVDR [24], where both of these beamformers are based on the constrained minimization of the beamformer output power. However, the BLCMV is derived under multiple linear constraints, including a unity gain constraint in the target signal direction, which is also used in the BMVDR. In the BLCMV, having multiple constraints means that small gains are specified in directions corresponding to interferer sources. The left and right beamformer coefficients can be derived by the following constrained minimizations in (12) and (13), respectively. For simplicity, the $f$ and $t$ index are omitted here.

$$
\min_{\mathbf{w}} \mathbf{w}_l^H(\mathbf{R}_y)\mathbf{w}_l \text{ subject to } \mathbf{C}^H\mathbf{w}_l = \mathbf{g}_l
\tag{12}
$$

$$
\min_{\mathbf{w}} \mathbf{w}_r^H(\mathbf{R}_y)\mathbf{w}_r \text{ subject to } \mathbf{C}^H\mathbf{w}_r = \mathbf{g}_r
\tag{13}
$$

The constraint matrix $\mathbf{C}$ includes the directivity vectors (HRTFs) of each constraint direction, i.e., $\mathbf{C} = [\mathbf{d}(f,\theta_x), \mathbf{d}(f,\theta_{v1}), ..., \mathbf{d}(f,\theta_{vk})]$. The left gain vector is $\mathbf{g}_l = [\varsigma d_{1,l}(f,\theta_x), \eta d_{1,l}(f,\theta_{v1}), ..., \eta d_{1,l}(f,\theta_{vk})]^T$ and the right gain vector is $\mathbf{g}_r = [\varsigma d_{M/2+1,r}(f,\theta_x), \eta d_{M/2+1,r}(f,\theta_{v1}), ..., \eta d_{M/2+1,r}(f,\theta_{vk})]^T$. The scalars $\varsigma$ and $\eta$ should be in the range between 0 and 1. In order to guarantee the near distortionless response of the target, $\varsigma$ should be close to 1. The value of $\eta$ controls the noise reduction level. The number of constraints $k$ available for the interferers depends on the number of available microphones, such that $k \le M - 2$. In this work, as the 2 + 2 microphone configuration is used, $k \le 2$. In other words, assuming no DOA mismatch, the BLCMV [13], [14] can preserve the binaural cues for only two directional interferers when the 2 + 2 microphone configuration is used.

Using the complex Lagrangian multiplier method to solve the constrained optimization problems in (12) and (13), the left and right binaural beamformer coefficients $\mathbf{w}_l$ and $\mathbf{w}_r$ are as in (14) and (15), respectively:

$$
\mathbf{w}_l = \mathbf{R}_y^{-1}\mathbf{C}(\mathbf{C}^H\mathbf{R}_y^{-1}\mathbf{C})^{-1}\mathbf{g}_l
\tag{14}
$$

$$
\mathbf{w}_r = \mathbf{R}_y^{-1}\mathbf{C}(\mathbf{C}^H\mathbf{R}_y^{-1}\mathbf{C})^{-1}\mathbf{g}_r.
\tag{15}
$$

Note that some level of diagonal loading may be required in practice, to regularize the matrix inversions [46]. Different options for the choice of correlation matrices have previously been introduced for the BLCMV [32]. In (14) and (15), the simplest option from [32] which uses the noisy microphone signals correlation matrix $\mathbf{R}_y$ is considered. By using the noisy microphone signals correlation matrix $\mathbf{R}_y$, there is no need for a sophisticated target voice activity detector (VAD) to estimate the noise components correlation matrices $\mathbf{R}_v$ and $\mathbf{R}_n$. The two other suggested options in [32] are using either the overall noise components correlation matrix $(\mathbf{R}_v + \mathbf{R}_n)$ or the background diffuse-like noise correlation matrix $\mathbf{R}_n$. Using $\mathbf{R}_v + \mathbf{R}_n$ or $\mathbf{R}_n$ in the beamformer coefficients computation increases the robustness to mismatch between the estimated target directivity vector and the actual target directivity vector [47], because using the noise components correlation matrix (either $\mathbf{R}_v + \mathbf{R}_n$ or $\mathbf{R}_n$) in the minimization criteria of (12) and (13) does not lead to target components minimization. At the opposite, distortion/attenuation of the target component in the beamformer output signal can occur in the presence of mismatch if $\mathbf{R}_y$ is used (since $\mathbf{R}_y$ includes the target component).

However, estimating $\mathbf{R}_v$ is often a difficult task in nonstationary multiple talkers conditions. And even though $\mathbf{R}_n$ can be more easily estimated, for beamformers that do not rely on constraints at interferer directions to reduce the interferers (such as the BMVDR or our proposed method, as we will explain later), using $\mathbf{R}_n$ leads to a solution that is not capable of significantly reducing the interferers. On the other hand, using $\mathbf{R}_n$ in a beamformer such as the BLCMV [31], [32] can be sufficient as long as there are constraints in the interferers directions, since the reduction of interferers is then determined by the value of a small constraint gain $\eta$.

### B. Binaural MVDR (BMVDR) and Its BMVDR-n Extension

The BMVDR [48] is a special case of the BLCMV, with a single constraint in the estimated target direction. Therefore, the constraint matrix $\mathbf{C}$ can be reduced to $\mathbf{d}(f, \theta_x)$, and the gain vectors $\mathbf{g}_l$ and $\mathbf{g}_r$ can be reduced to $d_{1,l}(f, \theta_x)$ and $d_{M/2+1,r}(f, \theta_x)$, respectively. The BMVDR preserves the binaural cues of the target in case of no target DOA mismatch; however, it distorts the binaural cues for the directional interferers and the background noise. As an attempt to enhance the binaural cues preservation ability of the BMVDR for the noise components, a small portion of the original noisy signal can be added to the BMVDR, leading to the BMVDR with partial noise estimation (BMVDR-n). The idea of adding a small portion of the original noisy signal to the processed output was introduced in [29]. More details of the BMVDR and the BMVDR-n can be found in [44], [45]. Many extensions to the BMVDR beamformer were previously introduced in the literature, such as the work in [24]. However, in this work we will compare our proposed algorithm with the BMVDR-n, because of its ability to preserve the binaural cues for both the directional interferers and the diffuse-like background noise. For a fair comparison of our proposed algorithm with the BMVDR and BMVDR-n algorithms, the noisy correlation matrix $\mathbf{R}_y$ will be used as for the BLCMV.

## IV. THE PROPOSED BEAMFORMING ALGORITHM

In this section, a binaural TLCMV robust to target DOA mismatch is first introduced, which does not require estimates of the interferers' DOAs or propagation models. It should be noted that in some previous work such as in [31]–[36], the name BLCMV is used for beamformers that use multiple constraints in order to attenuate directional interferers (in addition to the constraint to preserve the target). However, in this work, we use the name TLCMV (Target-LCMV) for beamformers that use more than one (normally two) constraints for the target, and no constraint for the interferers. The proposed Robust TLCMV requires an estimate of the target DOA, but the true target DOA can be within $\pm 10$ degrees of the estimated target DOA, as will be shown through experiments. This is a realistic condition, however the actual estimation of the target DOA is not considered in this paper.

Since the Robust TLCMV distorts the binaural cues for the directional interferers and the background diffuse noise, a postprocessor which does not require directivity vectors information is also proposed in this section, to provide a good level of binaural cues preservation while providing good overall noise reduction.

### A. The Proposed Robust TLCMV Beamforming Algorithm

Aiming to design a binaural beamformer that provides little suppression for sources from angles within a small angular region around the estimated target direction, the Robust TLCMV is introduced. Two constraints with unity gains are used in the middle of each side of a target zone, which consists of $\pm 10$ degrees around the estimated target DOA. For example, if the estimated target direction is at 0 degree, the beamformer assumes that the target can be anywhere between $-10$ to 10 degrees, and two unity constraints are used at $\pm 5$ degrees, in the middle of each side in the estimated target zone. The constraints of the Robust TLCMV are as described in (16) and (17), with the beamformer coefficients computed as in (14) and (15):

$$\mathbf{C}^H \mathbf{w}_l = \mathbf{g}_l \quad \mathbf{C} = [\mathbf{d}(f, \theta_x + \Delta), \mathbf{d}(f, \theta_x - \Delta)]$$
$$\mathbf{g}_l = [d_{1,l}(f, \theta_x + \Delta), d_{1,l}(f, \theta_x - \Delta)] \tag{16}$$

$$\mathbf{C}^H \mathbf{w}_r = \mathbf{g}_r \quad \mathbf{C} = [\mathbf{d}(f, \theta_x + \Delta), \mathbf{d}(f, \theta_x - \Delta)]$$
$$\mathbf{g}_r = [d_{M/2+1,r}(f, \theta_x + \Delta), d_{M/2+1,r}(f, \theta_x - \Delta)] \tag{17}$$

where $\theta_x \pm \Delta$ are the directions of unity constraints in the middle of the assumed target zone. The gain values used in $\mathbf{g}_l$ and $\mathbf{g}_r$ ensure that the beamformer output for a source from DOAs $\theta_x + \Delta$ and $\theta_x - \Delta$ has the same level as the one found at the input reference microphone for that same source, which we will refer to as a "unit gain" (i.e., the gain is relative to the input reference microphone level). Using two unity constraints around the estimated target direction forces the beamformer to have a wider beam in the direction of the target. Figs. 2 and 3 illustrate beampatterns of a fixed BMVDR beamformer with a single constraint at 0 degree under 2-D (cylindrically isotropic) diffuse noise conditions and the beampatterns of a fixed Robust TLCMV beamformer with constraints at $+5$ and $-5$ degrees under 2-D diffuse noise conditions at different frequencies for the left and right side. The beampatterns are obtained with HRTFs measured from behind-the-ear (BTE) hearing aid units on a mannequin in an anechoic environment, using four microphone signals, i.e., 2 microphones at each ear. The same HRTFs are also used to produce the 2-D diffuse noise correlation matrix required to produce Figs. 2 and 3. The beampattern $BP_i(\theta)$ is computed as in (18):

$$BP_i(\theta) = |\mathbf{w}_i^H(f)\mathbf{d}(f,\theta)|^2 \qquad (18)$$

where $\mathbf{w}_i$ is the left binaural beamformer coefficients $\mathbf{w}_l$ or the right binaural beamformer coefficients $\mathbf{w}_r$. Figs. 2 and 3 show that for higher frequencies the BMVDR has a narrow beam around the target direction, i.e., 0 degree. This narrow beam around the target direction indicates that the BMVDR is not robust to small target DOA mismatch. However, the Robust TLCMV has a wider beam around the target direction over all frequency components, therefore by design it is more robust to target DOA mismatch. There is a trade-off between robustness to target DOA mismatch and noise reduction. The use of an additional constraint for the target in the TLCMV design leads to a reduction in the degrees of freedom available for noise reduction (e.g., the positioning of adaptive "nulls"). However, due to the sparsity and the disjoint properties of speech signals in practice [49], there are often only one or sometimes two dominant directional interferer sources active at each time-frequency bin, and having two degrees of freedom left for the beamformer can be sufficient for good adaptive noise reduction, as will be illustrated later in this paper.

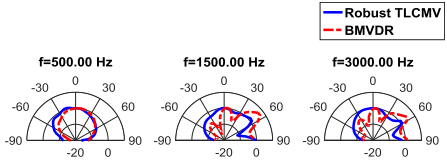

Fig. 2. Beampatterns of BMVDR and Robust TLCMV at different frequencies, shown for left side.

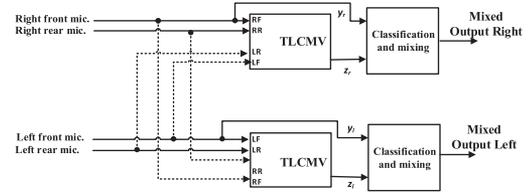

Fig. 4. The Robust TLCMV with CCMBB post-processor (dotted lines represent signals transmitted through a wireless link).

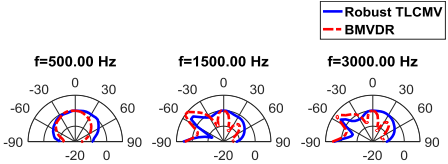

Fig. 3. Beampatterns of BMVDR and Robust TLCMV at different frequencies, shown for right side.

### B. Post-Processor Using Modified Coherence-Based Classification and Mixing Binaural Beamforming (CCMBB)

The proposed Robust TLCMV of the previous section can distort the binaural cues for the directional interferers and the diffuse-like background noise. In order to achieve better binaural cues preservations for these interferers and for diffuse noise components, while at the same time achieving a good level of overall reduction for interferers and diffuse noise, a post-processor based on time-frequency (T-F) classification and mixing of binaural signals is proposed in this section, as Fig. 4 shows. It is an updated version of our recent work [41], to provide a simpler and improved classification and mixing algorithm.

A complex coherence is computed for classification, as it gives the ability to exploit two classification decisions: one for the magnitude and one for the phase. We will thus refer to the post-processing algorithm as the Coherence-based Classification and Mixing for Binaural Beamforming (CCMBB). The complex coherence is computed on each side, between two signals locally available on each side. The first signal is the binaural beamformer outputs ($z_l(f,t)$ or $z_r(f,t)$, depending on the side), with a good level of interferers and diffuse noise reduction. The second signal is the front microphone noisy signal ($y_l(f,t)$ or $y_r(f,t)$), which fully preserves the binaural cues for all acoustic scene components. Alternatively, at the cost of increased complexity, the second signal could be a signal with an intermediate level of interferers and diffuse noise reduction but with binaural cues still preserved, such as the output from a common gain beamforming approach (e.g., [43], without the post-processing).

The left complex coherence $C_{zl,yl}(f,t)$ and right complex coherence $C_{zr,yr}(f,t)$ are computed as in (19) and (20), respectively:

$$C_{zl,yl}(f,t) = \frac{\Gamma_{zl,yl}(f,t)}{\sqrt{\Gamma_{yl,yl}(f,t)\Gamma_{zl,zl}(f,t)}} \qquad (19)$$

$$C_{zr,yr}(f,t) = \frac{\Gamma_{zr,yr}(f,t)}{\sqrt{\Gamma_{yr,yr}(f,t)\Gamma_{zr,zr}(f,t)}} \qquad (20)$$

where, $\Gamma_{yl,yl} = E\{|y_l(f,t)|^2\}$, $\Gamma_{yr,yr} = E\{|y_r(f,t)|^2\}$, $\Gamma_{zl,zl} = E\{|z_l(f,t)|^2\}$, $\Gamma_{zr,zr} = E\{|z_r(f,t)|^2\}$ are, respectively, auto-power spectral densities (auto-PSDs) for the front microphone noisy signals and the binaural beamformer outputs, and $\Gamma_{zl,yl} = E\{|z_l(f,t)y_l^*(f,t)|\}$, $\Gamma_{zr,yr} = E\{|z_r(f,t)y_r^*(f,t)|\}$ are cross-PSDs between the binaural beamformer outputs and the front microphone noisy signals.

For binaural cues preservation of a directional source, at low frequency components with wavelengths longer than the diameter of the head the interaural phase difference (IPD, defined in the next section) is more important than the interaural level difference (ILD, also defined in the next section) [50]. On the other hand the ILD is more important for high frequencies with wavelength components smaller than the head diameter, i.e., for frequencies higher than 1500 Hz. In the proposed CCMBB, on each side for low frequency components ($<1500$ Hz) the magnitude of the binaural output is simply the magnitude of the beamformer output (no mixing, no classification). This is because the output magnitude does not play a role in preserving the phase-based IPD binaural cues of the interferers (important at low frequencies), and the magnitude-based ILD is not important at low frequencies. Therefore, the magnitude of the binaural output at low frequencies keeps the emphasis on interferers/noise reduction. Similarly, in the proposed CCMBB, on each side for high frequency components ($>1500$ Hz) the phase of the binaural output is simply the phase of the beamformer output (no mixing, no classification). This is because the output phase does not play a role in preserving the magnitude-based ILD binaural cues of the interferers (important at high frequencies), and the phase-based IPD is not important at high frequencies. Therefore, the phase of the binaural output at high frequencies keeps the emphasis on interferers/noise reduction.

Another type of binaural cues will be considered in this work, for the preservation of the spatial impression of background diffuse noise: the Magnitude Squared Coherence (MSC, defined in the next section). The above processing implies that in the proposed CCMBB the magnitude information of binaural output signals is considered to be less important for preservation of MSC at low frequencies, and that the phase information of binaural output signals is considered to be less important for preservation of MSC at high frequencies.

Therefore, two classification and mixing systems need to be developed based on the complex coherence: one for the binaural output signal phase at low frequencies, and one for the binaural output signal magnitude at high frequencies. To better explain the rationale for the phase and magnitude classification performed at each T-F, a few additional equations are provided below. These equations are not required in the actual implementation of the CCMBB post-processor, unlike (19), (20). For simplicity, the left $l$ and right $r$ indices are dropped in these equations since the same equation applies to each side, and the time (frame) and frequency indices are also dropped. As before, $x_{ref}$ represents the target component at the reference microphone, and we define $u_{ref} = v_{ref} + n_{ref}$ as the sum of the directional interferers components $v_{ref}$ and the diffuse noise components $n_{ref}$ at the reference microphone. The corresponding components in the beamformer output signal are written as $z_x$ and $z_u$. Therefore, we have $y_{ref} = x_{ref} + u_{ref}$ as the noisy input signal at the reference microphone, and $z = z_x + z_u$ as the beamformer output, on each side and for each time and frequency bin.

Considering $z$ and $y_{ref}$ as zero-mean random variables and using the polar notation for these variables, the complex coherence becomes as follows, where $E\{.\}$ refers to an averaging process over consecutive frames in each frequency bin:

$$C_{z,y} = \frac{E\{z y_{ref}^*\}}{\sqrt{E\{|z|^2\}}\sqrt{E\{|y_{ref}|^2\}}}$$

$$= \frac{E\{|z_x||x_{ref}|e^{j(\sphericalangle z_x - \sphericalangle x_{ref})} + |z_u||u_{ref}|e^{j(\sphericalangle z_u - \sphericalangle u_{ref})}\}}{\sqrt{E\{|z_x|^2 + |z_u|^2\}}\sqrt{E\{|x_{ref}|^2 + |u_{ref}|^2\}}}. \tag{21}$$

The last part of (21) assumes that components from the target signal $x_{ref}$ and components from the "interferers plus diffuse noise" signal $u_{ref}$ are uncorrelated (as stated in a previous section). Next, if a target distortionless response is assumed for the beamformer, i.e., $z_x = x_{ref}$, (21) becomes:

$$C_{z,y} = \frac{E\{[|x_{ref}|^2 + |z_u|]|u_{ref}|e^{j(\sphericalangle z_u - \sphericalangle u_{ref})}\}}{\sqrt{E\{|x_{ref}|^2 + |z_u|^2\}}\sqrt{E\{|x_{ref}|^2 + |u_{ref}|^2\}}}. \tag{22}$$

At low frequencies, a larger phase change $|\sphericalangle z_u - \sphericalangle u_{ref}|$ between the input and output interferers/noise components is more likely to lead to distortion of interferers/noise IPD binaural cues between the left and right binaural outputs, because such changes do not occur symmetrically in the beamformer on each side of a binaural system. Similarly, at high frequencies a larger magnitude change $||z_u| - |u_{ref}||$ between the input and output interferers/noise components (i.e., a larger interferers/noise reduction) is more likely to lead to distortion of interferers/noise ILD binaural cues between the left and right binaural outputs. Evaluating from (22) the impact on $C_{z,y}$ of different $|\sphericalangle z_u - \sphericalangle u_{ref}|$ phase changes and different interferers/noise reduction levels, we can then use $C_{z,y}$ as a classification criterion for the CCMBB binaural output phase at low frequencies, where IPD is important. Likewise, evaluating from (22) the impact on $C_{z,y}$ of different $||z_u| - |u_{ref}||$ magnitude changes and different interferers/noise reduction levels, we can then use $C_{z,y}$ as a classification criterion for the CCMBB binaural output magnitude at high frequencies, where ILD is important.

First, we consider the effect of the phase change $|\sphericalangle z_u - \sphericalangle u_{ref}|$ for some important cases. The effect is more directly observed on the coherence phase value $|\sphericalangle C_{z,y}|$. From the numerator of (22), we see that a small coherence phase value $|\sphericalangle C_{z,y}|$ occurs if there is a small phase change $|\sphericalangle z_u - \sphericalangle u_{ref}|$ (regardless of the interferers/noise reduction level, i.e., level of $|z_u|$ relative to $|u_{ref}|$ and $|x_{ref}|$). Another case where a small coherence phase value $|\sphericalangle C_{z,y}|$ occurs is when there is a large $|\sphericalangle z_u - \sphericalangle u_{ref}|$ phase change with a strong interferers/noise reduction ($|z_u|$ small relative to $|u_{ref}|$ and $|x_{ref}|$). A case producing a large coherence phase value $|\sphericalangle C_{z,y}|$ is when a large $|\sphericalangle z_u - \sphericalangle u_{ref}|$ phase change is combined with weak interferers/noise reduction ($|z_u|$ level similar to $|u_{ref}|$ and $|x_{ref}|$ levels).

Since the case with a large coherence phase value $|\sphericalangle C_{z,y}|$ mentioned above includes both weak interferers/noise reduction and increased risk of binaural IPD cues distortion (from the large $|\sphericalangle z_u - \sphericalangle u_{ref}|$ phase change), the CCMBB does not use the beamformer output phase in such case. However, to avoid losing cases with good interferers/noise reduction levels, the CCMBB keeps the beamformer output phase for smaller values of $|\sphericalangle C_{z,y}|$ (which includes some cases with good or weak amount of interferers/noise reduction, as well as large or small $|\sphericalangle z_u - \sphericalangle u_{ref}|$). The resulting set of equations for the CCMBB binaural output phase component at low frequencies is:

$$\sphericalangle(z_{m,l}(f,t)) = \begin{cases} \sphericalangle(y_l(f,t)), & |\sphericalangle(C_{zl,yl}(f,t))| > \mu\pi \\ \sphericalangle(z_l(f,t)), & |\sphericalangle(C_{zl,yl}(f,t))| \le \mu\pi \end{cases} \tag{23}$$

$$\sphericalangle(z_{m,r}(f,t)) = \begin{cases} \sphericalangle(y_r(f,t)), & |\sphericalangle(C_{zr,yr}(f,t))| > \mu\pi \\ \sphericalangle(z_r(f,t)), & |\sphericalangle(C_{zr,yr}(f,t))| \le \mu\pi \end{cases} \tag{24}$$

The threshold value is a tunable parameter $\mu$ $\pi$ ($0 < \mu < 1$), where a lower $\mu$ leads to lower IPD binaural cues errors (and lower MSC errors), but also to lower interferers and diffuse noise reduction. A value of $\mu = 0.1$ has been found to provide satisfactory experimental results in our simulations.

Next, we consider the effect of the magnitude change $||z_u| - |u_{ref}||$ for some important cases. The effect is more directly observed on the coherence magnitude value $|C_{z,y}|$. From (22), we see that a case producing a smaller coherence magnitude value $|C_{z,y}|$ is when there is good interferers/noise reduction performance (small $|z_u|$ level relative to $|u_{ref}|$ and $|x_{ref}|$, and therefore large $||z_u| - |u_{ref}||$). On the other hand, if $||z_u| - |u_{ref}||$ is small (weak interferers/noise reduction, $|z_u|$ level similar to $|u_{ref}|$ and $|x_{ref}|$ levels), the value of $|C_{z,y}|$ depends on the $|\sphericalangle z_u - \sphericalangle u_{ref}|$ phase change: if there is a large $|\sphericalangle z_u - \sphericalangle u_{ref}|$ phase change it leads to a smaller coherence magnitude value $|C_{z,y}|$, and if there is a small $|\sphericalangle z_u - \sphericalangle u_{ref}|$ phase change it leads to a larger coherence magnitude value $|C_{z,y}|$ (closer to 1.0).

We note that unlike the low frequency classification with coherence phase value $|\sphericalangle C_{z,y}|$ considered earlier, here there is no case which has both a weak interferers/noise reduction and an increased risk of binaural cues distortion (i.e., a higher risk of binaural ILD cues distortion from a large magnitude change $||z_u| - |u_{ref}||$). This is because by definition $||z_u| - |u_{ref}||$ is indicative at the same time of the interferers/noise reduction level (a larger value of $||z_u| - |u_{ref}||$ is better) and the risk of binaural ILD cues distortion (a smaller value of $||z_u| - |u_{ref}||$ is better). Therefore, the approach proposed for the CCMBB binaural output magnitude at high frequencies is less drastic or less binary than the previous approach for the CCMBB binaural output phase at low frequencies, and it involves mixing together the beamformer output magnitude and the noisy reference input magnitude. The resulting set of equations for the binaural output magnitude at high frequencies is (at each T-F bin):

$$|z_{m,l}(f,t)| = \begin{cases} \text{if } |C_{zl,yl}(f,t)| < T_l(f) \\ \quad \alpha |z_l(f,t)| + (1-\alpha)|y_l(f,t)| \\ \text{if } |C_{zl,yl}(f,t)| \geq T_l(f) \\ \quad (1-\alpha)|z_l(f,t)| + \alpha |y_l(f,t)| \end{cases} \quad (25)$$

$$|z_{m,r}(f,t)| = \begin{cases} \text{if } |C_{zr,yr}(f,t)| < T_r(f) \\ \quad \alpha |z_r(f,t)| + (1-\alpha)|y_r(f,t)| \\ \text{if } |C_{zr,yr}(f,t)| \geq T_r(f) \\ \quad (1-\alpha)|z_r(f,t)| + \alpha |y_r(f,t)| \end{cases}. \quad (26)$$

The mixing parameter $\alpha$ ($0 \leq \alpha \leq 1$) affects the trade-off between the level of interferers/noise reduction and the preservation of the binaural ILD cues. As described in an earlier paragraph, the case with a good level of interferers/noise reduction occurs for a smaller value of $|C_{z,y}|$, and to preserve this case the CCMBB selects the condition with $|C_{z,y}|$ lower than a threshold $T$ as the condition which puts more weight on interferers/noise reduction, i.e., more weight on the magnitude of the beamformer output. This is at the expense of increasing the risk of binaural

ILD cues distortion. To help the balance and keep the binaural ILD cues distortion at a reasonable level, for the alternate condition with $|C_{z,y}|$ higher than a threshold $T$ the CCMBB puts more weight on the preservation of the binaural ILD cues, i.e., more weight on the magnitude of the noisy reference input signal. Essentially this simply means using a value $\alpha > 0.5$ in (25), (26). This approach has been validated in our experiments using the objective metrics presented in the next section, where it was found that a value of $\alpha = 0.7$ provided satisfactory experimental results (good overall trade-off between interferers/noise reduction and ILD distortion).

The threshold values $T_l(f)$ and $T_r(f)$ in (25), (26) are computed by taking the magnitude of the complex coherences estimated at each frequency bin from 219 ms of signals (40 frames, with overlap). This is unlike the coherence functions in (19), (20), (23)–(26), which are estimated with a shorter total time of 59 ms (only 10 frames, with overlap). The total time close to 200 ms was selected so that the method with a threshold could be used in future work under dynamic conditions (e.g., with head movements and dynamic sources). Using the CCMBB algorithm as a post-processor for the proposed Robust TLCMV, we will refer to the resulting beamforming algorithm as the "Robust TL-CMV with CCMBB".

## V. PERFORMANCE MEASUREMENT

To evaluate the performance of the proposed algorithm and the state of the art BLCMV, BMVDR and BMVDR-n algorithms, several objective metrics are used in this work. First, to measure the ability of the binaural beamformers to preserve the binaural cues, the interaural information between the left and right side signals is required. Formally, the Interaural Transfer Function (ITF) is defined as the ratio of a directional source component from the left to the right ear [30]. For simplicity, the ITF, ILD and IPD metrics are developed below for the case of a single source, more specifically a single interferer source. In the case of several interferers, in this work we apply the same equations to an equivalent interferer signal which consists of the sum of all interferer signals. All the performance measurements in this section are frequency dependent metrics; however, the frequency index $f$ is omitted for simplicity. The input ITF for an interferer component can be computed as in (27), where $\Gamma_{(vref,r),(vref,l)}$ is the cross-PSD between the interferer component at the front left and front right reference microphones, and $\Gamma_{(vref,l),(vref,l)}$ is the auto-PSD of the interferer component at the front left reference microphone:

$$ITF_{in,v} = \frac{\Gamma_{(vref,r),(vref,l)}}{\Gamma_{(vref,l),(vref,l)}}. \quad (27)$$

Similarly, the ITF between the left and right beamformer outputs can be described by (28):

$$ITF_{out,v} = \frac{\Gamma_{(zv,r),(zv,l)}}{\Gamma_{(zv,l),(zv,l)}} \quad (28)$$

where $z_v$ is the interferer component in the beamformer output signals. The errors (or losses) in the Interaural Level Difference (ILD) and Interaural Phase Difference (IPD) binaural cues are defined as in (29) to (34):

$$ILD_{in,v} = 10 \log 10 |ITF_{in,v}|^2 \qquad (29)$$

$$ILD_{out,v} = 10 \log 10 |ITF_{out,v}|^2 \qquad (30)$$

$$\Delta ILD_v = ILD_{out,v} - ILD_{in,v} \qquad (31)$$

$$IPD_{in,v} = \sphericalangle ITF_{in,v} \qquad (32)$$

$$IPD_{out,v} = \sphericalangle ITF_{out,v} \qquad (33)$$

$$\Delta IPD_v = IPD_{out,v} - IPD_{in,v}. \qquad (34)$$

In this work, the ILD error $\Delta ILD_v$ is only computed for the frequency components above 1500 Hz, and the IPD error $\Delta IPD_v$ is only computed for the frequency components below 1500 Hz.

In order to preserve the spatial impression of the diffuse-like noise, the MSC of the binaural diffuse-like noise components also has to be preserved. The MSC between the reference microphones can be computed as in (35):

$$MSC_{n,in} = \left| \frac{\Gamma_{(nref,r),(nref,l)}}{\sqrt{\Gamma_{(nref,l),(nref,l)}\Gamma_{(nref,r),(nref,r)}}} \right|^2. \qquad (35)$$

where $\Gamma_{(nref,r),(nref,l)}$, $\Gamma_{(nref,l),(nref,l)}$ and $\Gamma_{(nref,r),(nref,r)}$ are cross- and auto-PSDs from the diffuse noise component at the front microphones.

Similarly, the MSC between the left and right binaural outputs can be computed as in (36):

$$MSC_{n,out} = \left| \frac{\Gamma_{(zn,r),(zn,l)}}{\sqrt{\Gamma_{(zn,l),(zn,l)}\Gamma_{(zn,r),(zn,r)}}} \right|^2 \qquad (36)$$

where $\Gamma_{(zn,r),(zn,l)}$, $\Gamma_{(zn,l),(zn,l)}$ and $\Gamma_{(zn,r),(zn,r)}$ are cross- and auto-PSDs from the diffuse noise component in the beamformer outputs. The MSC error is then computed as in (37):

$$\Delta MSC_n = MSC_{n,out} - MSC_{n,in}. \qquad (37)$$

Next, to measure the reduction of the interferers and diffuse noise components with the beamforming process, a signal to noise ratio gain (SNR-gain, array gain), a signal to interferers ratio gain (SIR-gain), and a signal to diffuse noise ratio gain (SDNR-gain) are computed on each side, providing the difference in dB between the SNR, SIR, and SDNR at the beamformer output and at the input reference microphone:

$$SNRgain(dB) = 10 \log \left( \frac{\Gamma_{zx,zx}}{\Gamma_{(zv+zn),(zv+zn)}} \right)$$
$$- 10 \log \left( \frac{\Gamma_{xref,xref}}{\Gamma_{(vref+nref),(vref+nref)}} \right) \qquad (38)$$

$$SIRgain(dB) = 10 \log \left( \frac{\Gamma_{zx,zx}}{\Gamma_{zv,zv}} \right) - 10 \log \left( \frac{\Gamma_{xref,xref}}{\Gamma_{vref,vref}} \right) \qquad (39)$$

$$SDNRgain(dB) = 10 \log \left( \frac{\Gamma_{zx,zx}}{\Gamma_{zn,zn}} \right) - 10 \log \left( \frac{\Gamma_{xref,xref}}{\Gamma_{nref,nref}} \right) \qquad (40)$$

where in the above cross- and auto-PSDs $x_{ref}$, $v_{ref}$ and $n_{ref}$ refer to the target, interferers and diffuse noise components at a reference microphone, while $z_x$, $z_v$ and $z_n$ refer to the corresponding components in the beamformer output signal.

Finally, to measure the target distortion on each side after processing, two measurements are used: a target Speech Distortion Ratio (SDR) and a Speech Distortion Magnitude-only distance (SDmag). For each side, we define a target distortion error signal $x_{dist}$ as the time domain difference between the (aligned) target component in the beamformer output $z_x$ and the target component at the reference microphone signal $x_{ref}$. The SDR is then computed with the auto-PSDs as in (41):

$$SDR = 10 \log \left( \frac{\Gamma_{xref,xref}}{\Gamma_{xdist,xdist}} \right), \qquad (41)$$

and the SDmag is computed with the same auto-PSDs but as in (42):

$$SDmag = |10 \log \Gamma_{xref,xref} - 10 \log \Gamma_{zx,zx}|. \qquad (42)$$

Since the computation of the performance metrics requires knowing the separate components in the beamformer output signals (target, interferers, diffuse noise), the so-called shadow-filtering method was used in the simulations, i.e., filtering/processing all the signal components individually with the same time-variant filter coefficients or post-filtering. In addition, since all the talker speech sources were always active in our simulations (except for normal short pauses between words), for each component all the computed frames were used to estimate the PSD statistics, and therefore no VAD was required under this setup.

## VI. EXPERIMENTAL SETUP

Head Related Transfer Functions (HRTFs) measured from a KEMAR mannequin wearing two binaural Behind-The-Ear (BTE) hearing aids are used for the simulations. The HRTFs were provided by a hearing aid manufacturer. There were two sets of HRTFs: HRTFs from an anechoic environment, and HRTFs from a mildly reverberant environment (T60 $\approx$ 150 ms). For our simulations, the directional signals (target, interferers) for the reverberant conditions are generated using the reverberant HRTFs. Beamformer designs are always performed using the anechoic HRTFs, and these HRTFs are also used to generate the directional signals for the subset of simulations with anechoic conditions. The distance used for the reverberant and the anechoic HRTFs measurements, which is between a loudspeaker source and the center of the head, was 1 m. The diffuse-like background noise recordings were also provided by a hearing aid manufacturer, again recorded on a KEMAR mannequin wearing two binaural BTE hearing aids, with babble noise recordings played at eight loudspeakers on a circle with a radius of 1 m around the KEMAR mannequin. The audio signals are sampled at 24 kHz. A Short Time Fourier Transform (STFT) is used to decompose the signals in the time-frequency domain, with a FFT size of 256 (10.67 ms), using a Hann window with 50% overlap between consecutive windows. The generated noisy mixtures of signals have a total length of 10 sec.

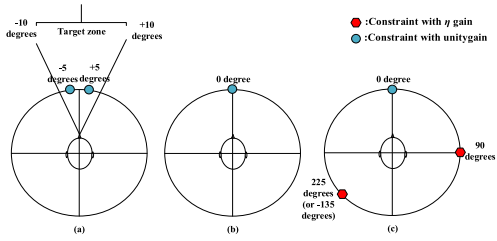

Fig. 5. The constraints directions for (a) proposed Robust TLCMV, (b) BMVDR and BMVDR-n (c) BLCMV.

## VII. SYSTEM EVALUATION AND SIMULATION RESULTS

In this section, the performance of our proposed beamformer "Robust TLCMV" is first compared with the BMVDR, which has more degrees of freedom available for noise reduction (more adaptive "nulls"), in order to assess the effect of reducing the number of degrees of freedom for the Robust TLCMV. In addition, the performance of the Robust TLCMV is evaluated using both noisy correlation matrix $\mathbf{R}_y$ and diffuse noise correlation matrix $\mathbf{R}_n$. Binaural cues preservations are not considered in these first comparisons. The proposed CCMBB post-processor for binaural cues preservation is then combined with the BMVDR and compared with the MVDR-n, to compare noise reduction, target distortion, and binaural cues preservation between these two approaches for cues preservation.

The proposed Robust TLCMV with CCMBB is then evaluated and compared with the BLCMV [13], [14]. For these algorithms, two types of propagation model mismatch are evaluated. The first type of mismatch is generated from the difference between the estimated and the true direction of arrivals for the directional sources, i.e., target and directional interferers. We will refer to this type of mismatch as DOA mismatch. The second type of mismatch is between the reverberant HRTFs used to generate the reverberant signals at the microphones and the anechoic HRTFs used in all the beamformer designs. We will refer to this second type of mismatch as HRTF mismatch.

For a frontal or near-frontal target case, the estimated target DOA is at 0 degree (for our proposed Robust TLCMV with and without CCMBB, and for the BMVDR, the BMVDR-n and the BLCMV) and the estimated interferers DOAs at 225 degrees and 90 degrees (with such estimates required for the BLCMV only). As our proposed Robust TLCMV beamformer design assumes that the true target DOA is within $\pm 10$ degrees of the estimated target DOA, two unity constraints are positioned in the middle of the estimated target zone at $\pm 5$ degrees as Fig. 5(a) illustrates, unlike the BMVDR and BMVDR-n which only use one constraint at the estimated target direction as Fig. 5(b) shows. On the other hand, the BLCMV uses three constraints: at 0 degree with gain $\zeta = 1$, and at 225 and 90 degrees with a gain $\eta$ set to 0.2 (as recommended in [32] and shown in Fig. 5(c)). A non-frontal target case with a target speaker at 90 degrees is also considered, with two unity constraints positioned in the middle of the estimated target zone, i.e., at $\pm 5$ degrees deviation from the assumed target direction in the Robust TLCMV, while the BLCMV again uses a unity gain constraint in the estimated target direction, and two constraints of gain $\eta$ at the estimated interferer directions.

| Acoustic scenarios | Target (°) | Interferers (°) | Diffuse noise |
|---|---|---|---|
| Scenario 1 | 0 or 10 | - | Same as the target level |
| Scenario 2 | 0 or 10 | 225 | No diffuse noise |
| Scenario 3 | 0 or 10 | 225,90 | No diffuse noise |
| Scenario 4 | 0 or 10 | 225,90 | Same as the target and interferers levels |

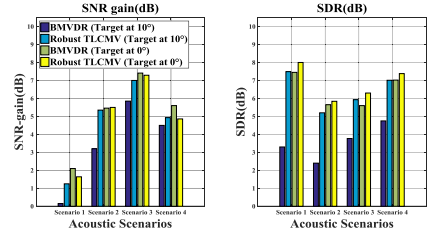

Fig. 6. Performance of BMVDR and Robust TLCMV in terms of SNR-gain and SDR, under acoustic scenarios from Table I (with and without target DOA mismatch).

### A. Robust TLCMV and BMVDR (Without Post-Processor)

In order to evaluate the performance of the proposed Robust TLCMV with the BMVDR (the first option in [24], which does not preserve the binaural cues of interfering sources), four different acoustic scenarios are used, each with a target at 0 or 10 degrees, as Table I illustrates. Due to space limitations, performance in this subsection is only shown in terms SNR-gain and SDR. The noise reduction and target distortion measurements in this section and the other sections are only shown for the "better ear" (the side where the input SNR is higher). The resulting performance metrics in Fig. 6 illustrate the effect of the target DOA mismatch in the performance of BMVDR and the proposed Robust TLCMV. The Robust TLCMV outperforms the BMVDR in terms of SDR under the four acoustic scenarios (more significantly for cases with DOA mismatch, i.e., target at 10 degrees), and it outperforms the BMVDR in terms of SNR-gain under the acoustic scenarios in the presence of DOA mismatch (target at 10 degrees). For acoustic scenarios with a target at 0 degree and no DOA mismatch, the proposed Robust TLCMV also slightly outperforms the BMVDR in terms of SDR. While this may seem surprising, it is because of HRTF mismatch (mismatch between anechoic HRTFs used to design the beamformer and reverberant HRTFs used to generate directional sources). Although it was designed for robustness to DOA mismatch, the Robust TLCMV with a wider beampattern around the estimated target direction is found to also provide better robustness to HRTF mismatch (here and in other results). In terms of noise reduction, for these ideal cases with no DOA mismatch the BMVDR outperforms the Robust TLCMV, although typically only by a fraction of a dB. Overall, the results show that the performance of the proposed Robust TLCMV is competitive (and significantly better in cases of DOA mismatch) compared to the BMVDR, despite a reduced number of degrees of freedom available for noise reduction.

The performance of the proposed Robust TLCMV is then evaluated using a noisy signals correlation matrix $\mathbf{R}_y$ as in (14)

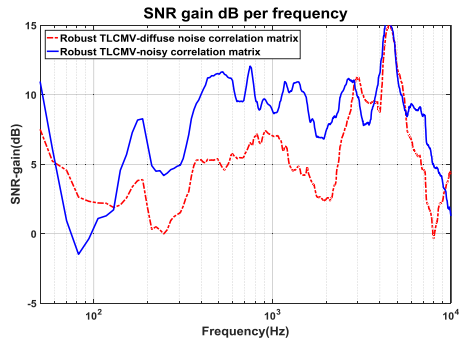

Fig. 7.  Performance of Robust TLCMV using noisy correlation matrix and diffuse noise correlation matrix in terms of SNR-gain.

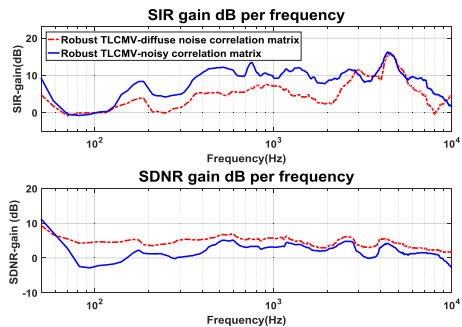

Fig. 8.  Performance of Robust TLCMV using noisy correlation matrix and diffuse noise correlation matrix in terms of SIR-gain and SDNR-gain.

and (15), and using a background diffuse-like noise correlation matrix $\mathbf{R}_n$ instead of $\mathbf{R}_y$ in (14) and (15). The $\mathbf{R}_n$ and $\mathbf{R}_y$ correlation matrices were estimated using a moving average lowpass first order recursive filter with a forgetting factor of 0.985. An acoustic scenario is used with a target at 0 degree, interferers at 225, 90 and 180 degrees, and diffuse-like noise (14 dB lower that the directional sources level). The resulting performance in terms of SNR-gain in Fig. 7 shows that using $\mathbf{R}_y$ for coefficients computation in the Robust TLCMV outperforms using $\mathbf{R}_n$. More detailed results are shown in Fig. 8 in terms of SIR-gain and SDNR-gain. The results illustrate the better performance of the proposed Robust TLCMV in terms of SIR-gain when $\mathbf{R}_y$ is used for the coefficients computation. This result can be justified since using $\mathbf{R}_y$ enables the proposed Robust TLCMV to adaptively position the nulls in the direction of the active interferers sources at each T-F bin. On the other hand, using $\mathbf{R}_n$ for coefficients computation in the Robust TLCMV performs better than using $\mathbf{R}_y$ for the SDNR-gain (diffuse noise reduction), which is normal since $\mathbf{R}_n$ is specifically tuned for that. In the rest of this paper, the noisy signals correlation matrix $\mathbf{R}_y$ is used in all simulations.

### B. CCMBB and a Method With Direct Mixing

In order to evaluate the performance of the proposed CCMBB post-processor for cues preservation separately from the proposed Robust TLCMV, the CCMBB is used as a post-processor to the BMVDR (BMVDR-CCMBB). The performance of the BMVDR-CCMBB is compared with the BMVDR (no binaural cues preservation for the interferers and noise components) and with a BMVDR-n which uses 0.7 of the beamformer output mixed with 0.3 of the noisy input signal. An acoustic scenario is used with a target at 0 degree (no DOA mismatch), an interferer at 165 degrees, and diffuse-like noise (5 dB below the directional sources level). The resulting performance metrics in Table II show that the proposed CCMBB cues preservation post-processing method combined with the BMVDR outperforms the BMVDR-n in terms of SNR-gain by around 2 dB, with a better SDmag distortion (2.3 dB) and similar scores for the other indicators. At the same time, the BMVDR-CCMBB has only a slightly lower SNR-gain than the BMVDR, while providing much better scores for the other metrics. This overall indicates the good performance of the CCMBB post-processor.

| Method | SNR-gain (dB) | SDR (dB) | SDmag (dB) | ILD-error (dB) | IPD-error (rad.) | MSC-error |
|---|---|---|---|---|---|---|
| BMVDR | 5.93 | 7.61 | 8.62 | 4.06 | 0.55 | 0.79 |
| BMVDR-n | 3.77 | 8.04 | 5.23 | 0.75 | 0.14 | 0.27 |
| BMVDR-CCMBB | 5.21 | 8.07 | 2.96 | 0.36 | 0.11 | 0.21 |

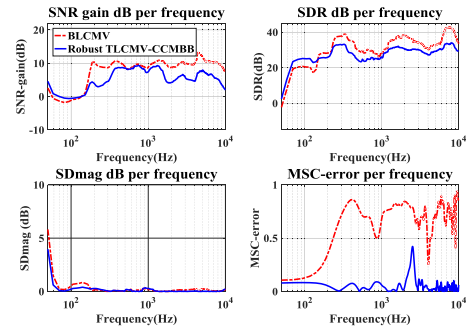

Fig. 9.  Performance in terms of SNR-gain, SDR, SDmag and MSC-error, with no DOA mismatch and no HRTF mismatch.

### C. Robust TLCMV With CCMBB and DOA Mismatch

In this section, the effect of the DOA mismatch for the target speaker as well as for the directional interferers is studied. We first evaluate the performance of the algorithms in an anechoic environment, using speech sources generated by anechoic HRTFs, in order to remove the other source of mismatch generated from the reverberation, i.e., HRTF mismatch.

To begin, a case with a target at 0 degree and interferers at 90 and 225 degrees is considered. For the BLCMV, this is an ideal case with no DOA mismatch, while for the proposed Robust TLCMV with CCMBB, the constraints set at ±5 degrees do not match the true target DOA (less ideal case). The target and the interferers all have the same level, and the diffuse noise level is set to 5 dB below each directional source level. In terms of SNR-gain, the resulting performance metric in the first plot of Fig. 9 illustrates the better SNR-gain performance of the BLCMV under this scenario ideal for it. In this scenario, the proposed Robust TLCMV with CCMBB does not have an exact unit constraint at 0 degree (true target DOA), unlike the BLCMV. Nevertheless, both the BLCMV and our proposed Robust TLCMV with CCMBB algorithm generate an output with significant SNR-gain and very low target distortion, as shown by the SNR-gain, SDR and SDmag plots of Fig. 9. Fig. 9 also clearly illustrates the effect of adding the CCMBB to preserve the spatial impression (binaural cues) of the diffuse-like background noise, as measured with the MSC-error metric. The BLCMV does not preserve the diffuse noise binaural cues, causing the large MSC-error scores.

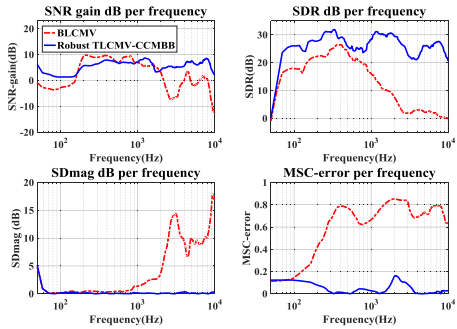

Fig. 10.    Performance in terms of SNR-gain, SDR, SDmag and MSC-error, under acoustic scenario with 10 degrees DOA mismatch and no HRTF mismatch.

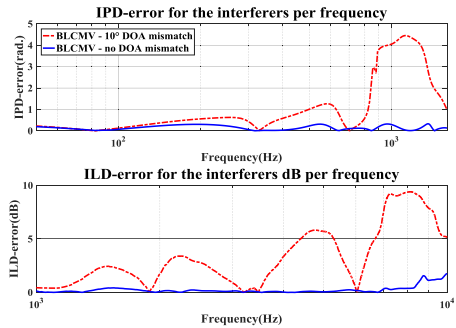

Fig. 11.    Performance of BLCMV in terms of IPD-error and ILD-error under anechoic acoustic scenario, without and with 10 degrees DOA mismatch.

Assuming an exact knowledge of the true DOA of the target as well as true DOAs of the directional interferers is impractical. Therefore, a case with 10 degrees of DOA mismatch is then tested, using an acoustic scenario with a target at 10 degrees, interferers at 235 and 100 degrees, and diffuse-like noise, all with the same levels as earlier. The resulting performance in terms of SNR-gain, SDR and SDmag in Fig. 10 illustrates that the Robust TLCMV with CCMBB provides significantly better results in this case with DOA mismatch, especially for high frequencies, i.e., above 1000 Hz. The post-processing CCMBB method again provides significant improvements in terms of diffuse-noise MSC-error. These results indicate the robustness of the proposed algorithm in the presence of target DOA mismatch. Moreover, since our proposed algorithm does not assume a prior knowledge of the directional interferers DOAs, its binaural cues preservation performance is not affected with interferers DOA mismatch, unlike the BLCMV. Fig. 11 shows that with 10 degrees DOA mismatch in an anechoic environment with a target at 10 degrees, interferers at 235 and 100 degrees, and diffuse-like noise, the abilities of the BLCMV to preserve the binaural cues of the directional interferers significantly decrease (i.e., increase in the IPD-error and ILD-error metrics).

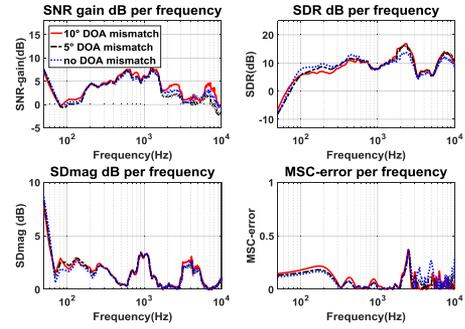

Fig. 12.    Performance of Robust TLCMV with CCMBB post-processor under mildly reverberant acoustic scenario (HRTF mismatch), with and without DOA mismatch for the target.

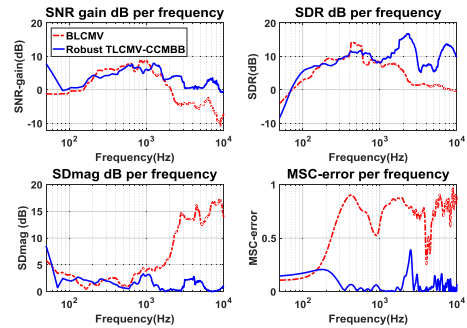

Fig. 13.    Performance in terms of SNR-gain, SDR, SDmag and MSC-error, under acoustic scenario with 10 degrees DOA mismatch for the target (and for interferers in the BLCMV), and HRTF mismatch.

### D. Robust TLCMV With CCMBB With DOA Mismatch and HRTF Mismatch

In this section, a more realistic evaluation is performed using speech signals generated in a mildly reverberant environment (T60 = approx. 150 ms). Three acoustic scenarios are generated with a target at 0, 5, or 10 degrees, interferers at 225 and 90 degrees, 230 and 95 degrees, or 235 and 100 degrees, as well as with diffuse noise. The target and the interferers again all have the same level, and the diffuse noise level is set to 5 dB below each directional source level. The directional signals were generated using reverberant HRTFs. The beamformer algorithms assume the same target DOA as before: 0 degree (for both algorithms), 90 and 225 degrees (required for BLCMV only). Therefore, these cases include HRTF mismatch, with and without DOA mismatch. The resulting performance metrics in Fig. 12 show that the proposed Robust TLCMV with CCMBB remains robust to DOA mismatch in the reverberant environment (up to 10 degrees) since it does not rely on constraints in the exact directions of the directional sources (unlike the BLCMV). Moreover, Fig. 13 illustrates the overall improved performance of the Robust TLCMV with CCMBB over the BLCMV in terms of SNR-gain, SDR, SDmag and MSC with 10 degrees DOA mismatch in the mildly reverberant environment, i.e., with HRTF mismatch. It is also noticeable that the BLCMV does not have the ability to preserve the spatial impression of diffuse noise in terms of MSC, unlike the proposed Robust TLCMV with CCMBB.

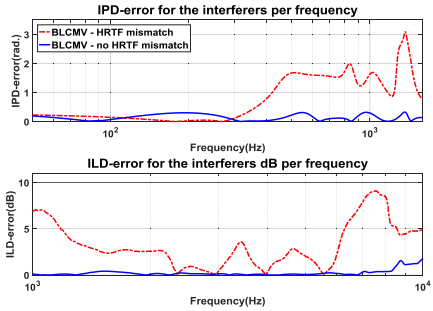

Fig. 14. Performance of BLCMV in terms of IPD-error and ILD-error with and without HRTF mismatch, and without DOA mismatch.

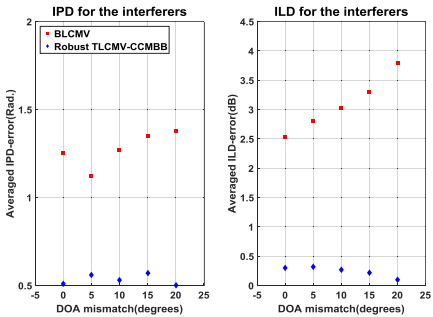

Fig. 15. Performance in terms of IPD-error and ILD-error under a reverberant acoustic scenario (HRTF mismatch), for different levels of interferer DOA mismatch.

To evaluate the effect of the HRTF mismatch separately, i.e., without the effect of DOA mismatch, on the ability of the BLCMV to preserve the binaural cues in terms of IPD and ILD, an acoustic scenario is generated with a target at 0 degree, interferers at 90 and 225 degrees, and diffuse noise (same levels as before). Fig. 14 shows that in reverberant environments, i.e., with HRTF mismatch, the ability of the BLCMV to preserve the binaural cues for the directional interferers significantly decreases. In order to evaluate the combined effect of the HRTF mismatch and DOA mismatch in the preservation of the binaural cues in terms of ILD and IPD, five acoustic scenarios are then used: acoustic scenarios without DOA mismatch, and with 5, 10, 15 and 20 degrees of interferers DOA mismatch. The resulting performance metrics in terms of IPD-error and ILD-error in Fig. 15 show the performance improvement of our proposed Robust TLCMV with CCMBB algorithm over the BLCMV for all the tested cases. The average IPD for the frequency components lower than 1500 Hz and the average ILD for the frequency components higher than 1500 Hz are shown in Fig. 15. For the case without interferer DOA mismatch, our proposed algorithm still outperforms the BLCMV in terms of IPD and ILD, because of the use of CCMBB post-processing. Fig. 15 also shows that our proposed Robust TLCMV with CCMBB is not affected by the increase in the interferer DOA mismatch combined with HRTF mismatch.

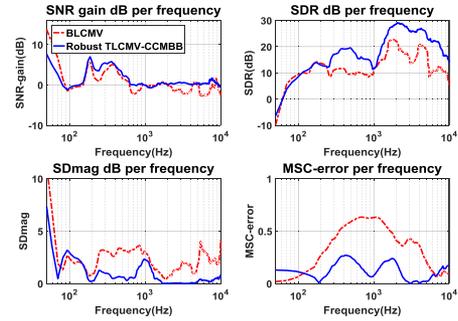

Fig. 16. For non-frontal target, performance in terms of SNR-gain, SDR, SD-mag and MSC-error under a reverberant acoustic scenario (HRTF mismatch), without DOA mismatch.

Further study of the HRTF mismatch effect is done under an acoustic scenario with a lateral target at 90 degrees, where the effect of the target HRTF mismatch can be more significant than for a frontal target case. Interferers at 225 and 315 degrees as well as diffuse noise are used. The target and the interferers all have the same level, and the diffuse noise level is set to 5 dB below each directional source level. The directional signals are generated using reverberant HRTFs, creating HRTF mismatch. In this case the beamformer algorithms know the value of the exact target DOA at 90 degree (for both algorithms, no target DOA mismatch), and the exact value of the interferers DOAs at 225 and 315 degrees (required for BLCMV only). Fig. 16 illustrates the improved performance of the proposed Robust TLCMV with CCMBB over the BLCMV in terms of noise reduction, target speech distortion, and preservation of the binaural spatial impression of the background diffuse noise for this scenario with HRTF mismatch.

## VIII. CONCLUSION

This work introduced a binaural beamforming algorithm robust to target DOA mismatch and HRTF mismatch (Robust TLCMV), as well as its combination with a post-processor to achieve a good trade-off between noise reduction and binaural cues preservation of all acoustic components (Robust TLCMV with CCMBB). The proposed robust beamformer does not require prior knowledge of the propagation model (e.g., HRTFs or HRTF ratios) for the directional interferers, or second order statistics estimation of the noise-only or interferers-only components. The Robust TLCMV was shown to produce better results than a BMVDR under the case of 10 degrees target DOA mismatch, and comparable performance for the ideal case of no target DOA mismatch. The CCMBB post-processor was shown to produce better results than a direct mixing of the beamformer output with the noisy input signal. Finally, the Robust TLCMV combined with the CCMBB was shown to produce better performance than the BLCMV under 10 degrees sources DOA mismatch and under HRTF mismatch with mild reverberation. Future work to further develop the proposed algorithm and validate its performance should include testing under environments with higher levels of reverberation, as well as with dynamic sources.

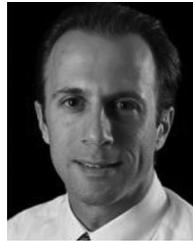

**Martin Bouchard** received the B.Ing., M.Sc.A., and Ph.D. degrees in electrical engineering from the Université de Sherbrooke, Sherbrooke, QC, Canada, in 1993, 1995, and 1997, respectively. In January 1998, he joined the School of Electrical Engineering and Computer Science, Faculty of Engineering, University of Ottawa, Ottawa, ON, Canada, where he is currently a Professor. In 1996, he co-founded SoftdB Inc., Quebec City, QC, which is still active today. Over the years, he has conducted research activities and consulting activities with more than 20 private sector and governmental partners, supervised more than 50 graduate students and postdoctoral fellows, and authored or coauthored more than 40 journal papers and 85 conference papers. His current research interests include signal processing methods in general and machine learning, with an emphasis on speech, audio, acoustics, hearing aids, and biomedical engineering applications. He served as a member of the Speech and Language Technical Committee of the IEEE Signal Processing Society from 2009 to 2011, as an Associate Editor for the *EURASIP Journal on Audio, Speech and Music Processing* from 2006 to 2011, and as an Associate Editor for the IEEE TRANSACTIONS ON NEURAL NETWORKS from 2008 to 2009. He is a member of the Ordre des Ingénieurs du Québec.

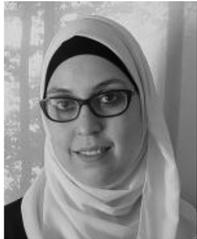

**Hala As'ad** received the M.A.Sc. degree in electrical engineering with a specialization in audio and speech processing in 2015 from the University of Ottawa, Ottawa, ON, Canada, where she is currently working toward the Ph.D. degree in electrical and computer engineering. Her doctoral research focuses on robust binaural beamforming, binaural cues preservation, and source direction of arrival detection in hearing aids. Her research interests include applied signal processing and machine learning with an emphasis on audio and speech processing, array signal processing, beamforming, speech enhancement, acoustic source localization, and hearing aids. She is the recipient of the Natural Sciences and Engineering Research Council Scholarship, the University of Ottawa Excellence Scholarship, and the Ontario Graduate Scholarship.

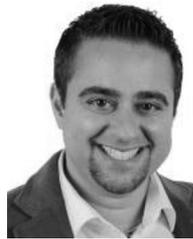

**Homayoun Kamkar-Parsi** received the B.A.Sc., M.A.Sc., and Ph.D. degrees in electrical engineering from the School of Information Technology and Engineering, University of Ottawa, Ottawa, ON, Canada, in 2001, 2004, and 2009, respectively. During his undergraduate studies, he has obtained the highest standing in his graduating class in Electrical Engineering and the Silver medal for the second Highest standing in entire Faculty of Engineering. His graduate scholarships included the Natural Sciences and Engineering Research Council scholarship and the Ontario Graduate Scholarship. Since 2009, he has been with Siemens Audiologische Technik GmbH (renamed as Sivantos GmbH in 2015 and as WS Audiology in 2019), Erlangen, Germany, where his main work and research include speech/audio signal processing with applications in speech enhancement, advanced multi-microphone beamforming for binaural hearing aids including remote external microphones (e.g., from smartphone), source localization and tracking, advanced scene analysis and machine learning (neural networks). In 2018, he was selected as one of the top inventors at Sivantos.